**Steps Towards a Capacitance Standard**

**Based on Single-Electron Counting at PTB**

H. Scherer, S. V. Lotkhov, G. D. Willenberg, and A. B. Zorin

Physikalisch-Technische Bundesanstalt Braunschweig

Bundesallee 100, 38116 Braunschweig, Germany

*Abstract* — **A capacitance standard based on the definition of capacitance $C = Ne/U$ is realized when a capacitor is charged with a known number $N$ of electrons ($e$ is the elementary charge) and the voltage $U$ across the capacitor is measured [1]. If $U$ is measured by means of a Josephson voltage standard, $C$ is solely determined by quantum numbers and fundamental constants. Presently, PTB is setting up this experiment using a special type of single-electron pump, the so-called r-pump [2], for transferring electrons one by one onto a cryogenic vacuum capacitor with the decadic capacitance $C = 1$ pF. We report on the progress in the setup with emphasis on the characterization of the single-electron tunneling elements. We describe the modifications of the single-electron circuit that are necessary to implement the capacitor charging experiment by using a four- or five-junction r-pump.**

---

The work was partially supported by the European Commission within the COUNT project.

## I. INTRODUCTION

Besides a suitable cryogenic capacitor ("cryocap", $C = 1$ pF), available at PTB [3], the key component of the experiment is the single-electron tunneling (SET) circuit. Since metrological impact is achieved only if the relative uncertainty in capacitance determination is better than $10^{-7}$, the accuracy of the charge transferred by the SET pump and controlled by an SET electrometer is crucial. The pioneering work at NIST has shown that this accuracy is achievable by using a seven-junction pump and demonstrated the first successful implementation of this so-called "Electron Counting Capacitance Standard" (ECCS) experiment [4]. In contrast to the NIST approach, which is complicated by the necessity to control six gate voltages, we aim at using so-called r-pumps, i.e. SET pumps with miniature series resistors on-chip, but with fewer junctions and gates, thus offering simplified operation [2].

## II. EXPERIMENTAL SETUP

A dilution refrigerator with an extended sample space diameter of 15 cm and a base temperature of about 10 mK was used for the experiments. Rf-filtering of ten signal lines (including up to four coaxial lines for the ac gate signals) was provided by two 1 m long pieces of a lossy miniature coaxial cable (CuNi, 0.3 mm diameter) fitted into each line, yielding an attenuation of about -120 dB at 10 GHz. The filtering cables are plugged onto a fixture on the mixing chamber. The construction permits us to easily enhance the filtering by adding 1 m long pieces of Thermocoax™ cable (0.34 mm diameter) to each line, bringing an additional attenuation of -190 dB at 10 GHz. A copper sample-holder box shields the SET circuit from ambient microwave irradiation and hosts the SET chip (wire-bonded into a chip carrier), the cryocap and two cryogenic needle switches. These switches, supplied by METAS within the COUNT project, are needed to connect the cryocap either to the SET pump (in order to charge it with single electrons), or to an ac capacitance bridge (for comparison with another capacitor). In an SET-box experiment the effective electron temperature was found to be $T_e \approx 100$ mK. For ac-driving of the SET pump gates, special electronic components (identical to the ones used at NIST [4]) were employed to generate triangular voltage pulses and to compensate parasitic cross-capacitances.

## III. Samples

Different SET circuits based on Al-$AlO_x$-Al tunnel junctions and on-chip Cr strip resistors were fabricated on thermally oxidized Si wafers using a three-angle evaporation technique [2]. Typical sample parameters of stand-alone three-junction (3j) r-pumps were $R_T = 65$ kΩ to 85 kΩ for the tunnel resistances, $C_T \approx 140$ aF to 160 aF for the junction capacitances, and $R = 50$ kΩ to 100 kΩ for the Cr strip resistances ($R > R_K = h/e^2 \approx 26$ kΩ). Cross-talk between gate electrodes and foreign islands was about 30%. Furthermore, two different types of samples were fabricated and tested with a 3j r-pump ($R \approx 50$ kΩ to 80 kΩ) connected to the input of an SET electrometer (Fig. 1). Type A samples were designed for single-electron trapping experiments. The small Al strip island between junction array and electrometer gate (5 µm long and 100 nm wide, total capacitance $C_\Sigma^m \approx 0.43$ fF) was intended to serve as a "memory node" for storing single charges. An additional memory gate electrode was coupled to that node via the capacitance $C_m \approx 65$ aF, and the coupling capacitance between memory node and electrometer island was $C_c \approx 20$ aF. The second sample type (B) was designed as a prototype for the final ECCS experiment. For this purpose, the input of an SET electrometer, serving as a null detector during the capacitor charging phase, was capacitively coupled to a common node with a 3j r-pump (interdigital coupler with $C_c \approx 0.8$ fF, Fig. 1). This node had the shape of a circular pad (110 µm diameter) and was planned to serve as a contact pad for the cryogenic needle switch.

## IV. MEASUREMENTS AND RESULTS

### A. *Three-Junction Stand-Alone R-Pumps*

Long-term measurements on 3j r-pumps generally revealed an excellent offset charge stability over several hours. After optimizing the gate voltage trajectory in the pumping mode (Fig. 2), i.e. driving the gates with ac pulses at cycle frequencies $f$ of few MHz, $I$-$V$ curves with current plateaus of $\approx 0.2$ mV width were observed at the quantized values $I = ef$ (Fig. 3, top). We performed a high-precision current measurement in the inflection point of the $I$-$V$ curve at $f = 2$ MHz, using a commercial electrometer with sub-fA current resolution [5] that we calibrated with a new technique [6]. Averaging 29 current measurements (made within 1 h), we found $I_{av} = 320.444$ fA with an expanded uncertainty of $\pm$ 0.033 fA, i.e. a relative uncertainty of $1 \cdot 10^{-4}$ (Fig. 3, bottom). Within this uncertainty we found no deviation from the quantized value $ef = 320.435$ fA.

### B. *"Type A" Samples (3j R-Traps with Electrometer)*

When $V_m$ was swept ($V_p$ grounded, see Fig. 1), charging of the memory node by electrons entering through the tunnel junction array was detected by the coupled electrometer. Tuning the electrostatic potentials of both islands by the gate voltages $V_1$ and $V_2$, the effective behavior of the array could be either set to the totally unblocked state, the "electron-box" state or the "trap" state. In the trap state both islands create an energy barrier that electrons have to surmount for entering or leaving the memory node. In the electrometer signal $I_{el}$ the resulting trapping of single-electrons on the memory node showed as typical hysteretic, steplike charging characteristic (Fig. 4). The width of the hysteresis loops could be varied and maximized by fine-tuning $V_1$ and $V_2$. Time series recordings of $I_{el}$ clearly showed "telegraph" switching events, reflecting changes between discrete charge states of the memory node. $V_m$ was balanced so that the average dwell times for two distinct memory node charge states became equal (close to the middle of a memory loop). In this state the heights of the energy barriers for both trapping and escape processes are equal ($\Delta U$), and the average dwell time was identified with the hold time $t_h$ of the junction array. For a three-junction trap with a series resistor of $R \approx 80$ kΩ we observed hold times of up to 16 s, which is slightly better than $t_h \approx 10$ s measured at NIST on a five-junction pump without resistor [7].

We measured the temperature dependence of $t_h$ in the range from $T = 25$ mK to 0.4 K and found a monotonic increase of the hold time when cooling the system from $T = 0.4$ K to 0.1 K (not shown). Below 0.1 K, $t_h$ seemed to be independent of the bath temperature, which reflects the mentioned temperature saturation in the electron system. From the Arrhenius plot of the $t_h(T)$ dependency a thermal activation energy ≈ 0.25 meV was derived which agrees well with $\Delta U$ estimated from an electrostatic model of the trap.

Although the r-trap design was not optimized for that purpose, we tried to operate the structure in the so-called "shuttle mode" [7], i.e. driving the pump gates such that one electron was quickly pumped onto the memory island (transfer time ≈ 190 ns) and pumped off after a wait time $\Delta t$. After optimizing $V_m$, time series records of the electrometer signal for $\Delta t = 0.5$ s, 1 s and 2 s clearly revealed discrete charging and discharging events that were statistically correlated to the wait time $\Delta t$, as indicated by the peaks in typical dwell time histograms (shown in Fig. 5 for $\Delta t = 1$ s). In the case of perfect shuttle-pumping this histogram would show a single peak at $\Delta t = 1$ s. The obvious background and the "revival" peaks at integer multiples of $\Delta t$, caused by "missed" shuttle cycles (dashed arrows in histogram plot), reflect rather frequent error processes occurring during shuttle-pumping. We attribute this to the small island capacitance $C_\Sigma^m \approx 0.43$ fF, causing a large potential change across the pump (≈ 0.4 meV) when a single electron is pumped on or off the island.

### *C. "Type B" Samples (3j R-Pumps Connected to Electrometer Via Needle Pad)*

The effective electrometer sensitivity (both signal/charge and signal/noise ratios) in type B samples was significantly suppressed due to the large capacitance of the interdigital gate, causing the total capacitance to be three times higher than for type A samples. Although the electrometer current showed gate modulation and the hold times are expected to be similar (in the order of 10 seconds) for type A and B samples, no single-electron charging events on the needle pad could be detected, for instance showing up as discrete jumps in time series measurements. We explain this by the large total capacitance $C_\Sigma^p \geq 30$ fF (estimated) of the pad island, causing a small "charge divider" ratio $C_c/C_\Sigma^p \leq 1/40$ with a correspondingly small fraction of the island charge appearing at the electrometer input. Despite this we could observe that the intensity of the electrometer signal fluctuations depended on the dc voltages ($V_1$, $V_2$) applied to the gates of the pump structure. A plot of the electrometer noise *vs.* $V_1$ and $V_2$ revealed a periodic square pattern (not shown) being congruent with the "honeycomb" patterns measured on stand-alone pumps (Fig. 2). Therefore we conclude that the observed charge fluctuations reflect the state of the pump, being either in an "open" state (triple point) or in a blockade state, i.e. inside a honeycomb cell.

## V. Conclusion and Outlook

On a zero-biased three-junction r-pump, current quantization according to $I = ef$ at $f = 2$ MHz was demonstrated by a direct current measurement with a relative uncertainty of $1 \cdot 10^{-4}$. The relatively easy operation of these pumps and their excellent stability in the pumping regime due to negligible offset charge changes suggests that they may be considered to serve as ultra-low-current sources for metrology purposes.

The hold time $t_h \approx 16$ s of a three-junction r-pump demonstrates the positive effect of the Cr strip resistors ($R \approx 80$ kΩ), but is still smaller than $t_h \approx 10$ min for the seven-junction pump in the NIST experiment [4]. Also, the temperature dependence of $t_h$ suggests that thermally activated processes limit the three-junction device performance down to the electron base temperature achieved in our setup. Therefore we will extend our investigations on four- and five-junction r-pumps to improve the electron transfer accuracy for the final capacitor charging experiment.

We investigated ECCS prototype structures but found no discrete single-electron charging events on the needle pad, which we assign to the insufficient charge resolution of the electrometer. A determination of the electron transfer accuracy, based on shuttle-pumping and hold time measurements on these samples, was thus not feasible. The poor sample performance was probably caused by the too large stray capacitance of the pad. Following [4], we are currently pursuing the sample preparation on quartz substrate ($\varepsilon \approx 4.5$) instead of silicon ($\varepsilon \approx 11$) and expect an increase in the charge divider ratio $C_c/C_\Sigma^p$ towards 1/20 (as in type A samples and in [4]). We believe that these improvements will enable us to implement the ECCS experiment with a four- or five-junction r-pump.

## ACKNOWLEDGEMENTS

We thank M. Busse for assistance with the cryogenic setup, U. Becker and H. N. Tauscher for help with the electronic equipment, F.-J. Ahlers for contributions to the measurement software, and Ch. Hof for fruitful discussions.

## References

[1] E. R. Williams, R. N. Ghosh, and J. M. Martinis, “Measuring the electron’s charge and the fine-structure constant by counting electrons on a capacitor,” *J. Res. Natl. Inst. Stand. Technol.*, vol. 97, pp. 299-304, Mar.-Apr. 1992.

[2] S. V. Lotkhov, S. A. Bogoslovsky, A. B. Zorin, and J. Niemeyer, “Operation of a three-junction single-electron pump with on-chip resistors,” *Appl. Phys. Lett.*, vol. 78, pp. 946-948, Feb. 2001.

[3] G.-D. Willenberg and P. Warnecke, “Stable cryogenic vacuum capacitor for single-electron charging experiments,” *IEEE Trans. Instr. Meas.*, vol. 50, pp. 235-237, Apr. 2001.

[4] M. W. Keller, A. L. Eichenberger, J. M. Martinis, and N. M. Zimmerman, “A capacitance standard based on counting electrons,” *Science*, vol. 10, pp. 1706-1709, Sept. 1999.

[5] Type UNIDOS E (Manufacturer PTW, Germany). Citation of the instrument type is for identification purposes and does not imply that it is the best available.

[6] G.-D. Willenberg, H.-N. Tauscher, and P. Warnecke, “A traceable precision current source for currents between 100 aA and 10 pA,” *IEEE Trans. Instr. Meas.*, vol. 52, pp. 436-439, Apr. 2003.

[7] J. M. Martinis, M. Nahum, and H. D. Jensen, “Metrological accuracy of the electron pump,” *Phys. Rev. Lett.*, vol. 72, pp. 904-907, Feb. 1994.

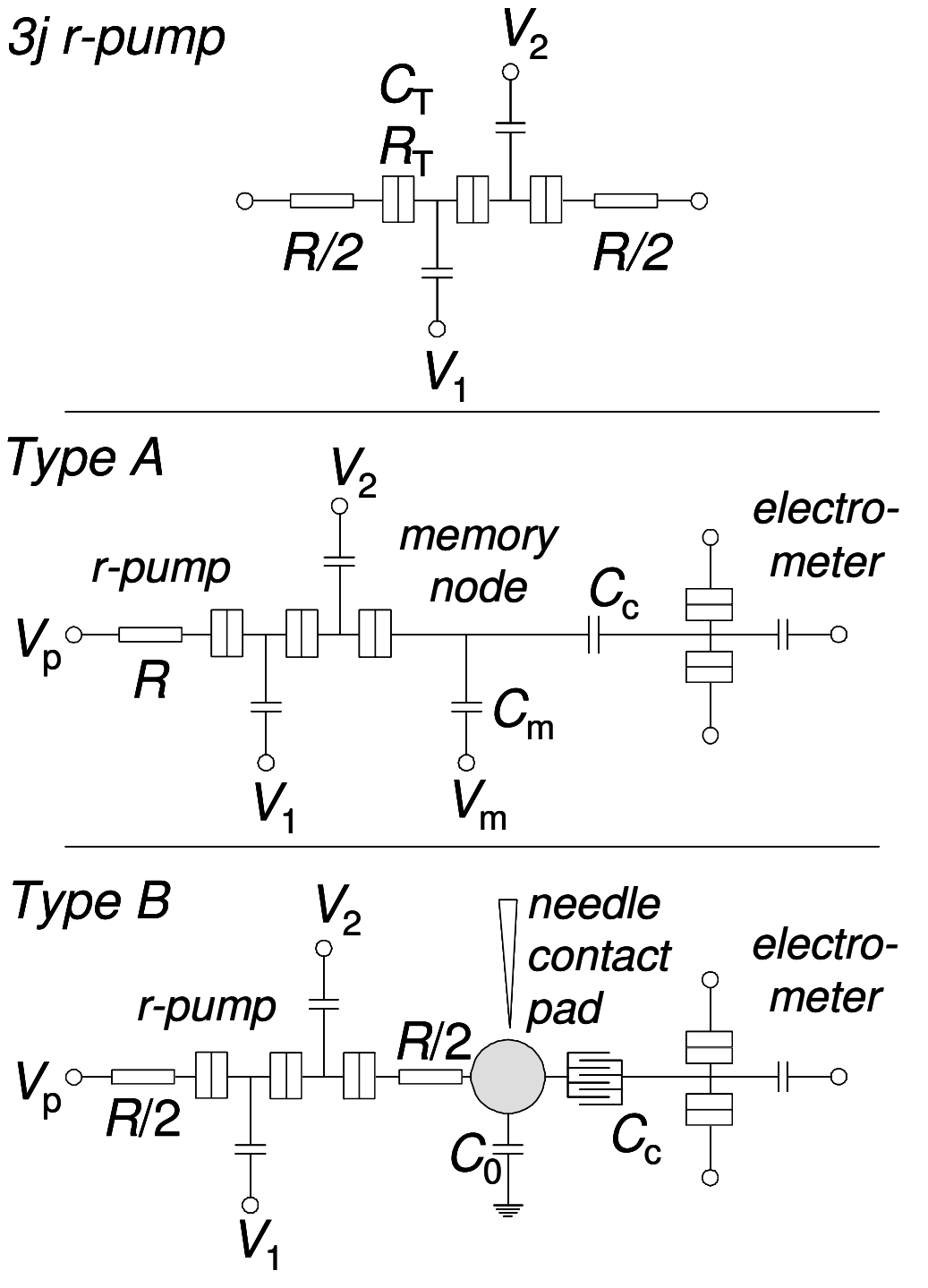

***Fig. 1: Circuit schemes of different sample types investigated.***

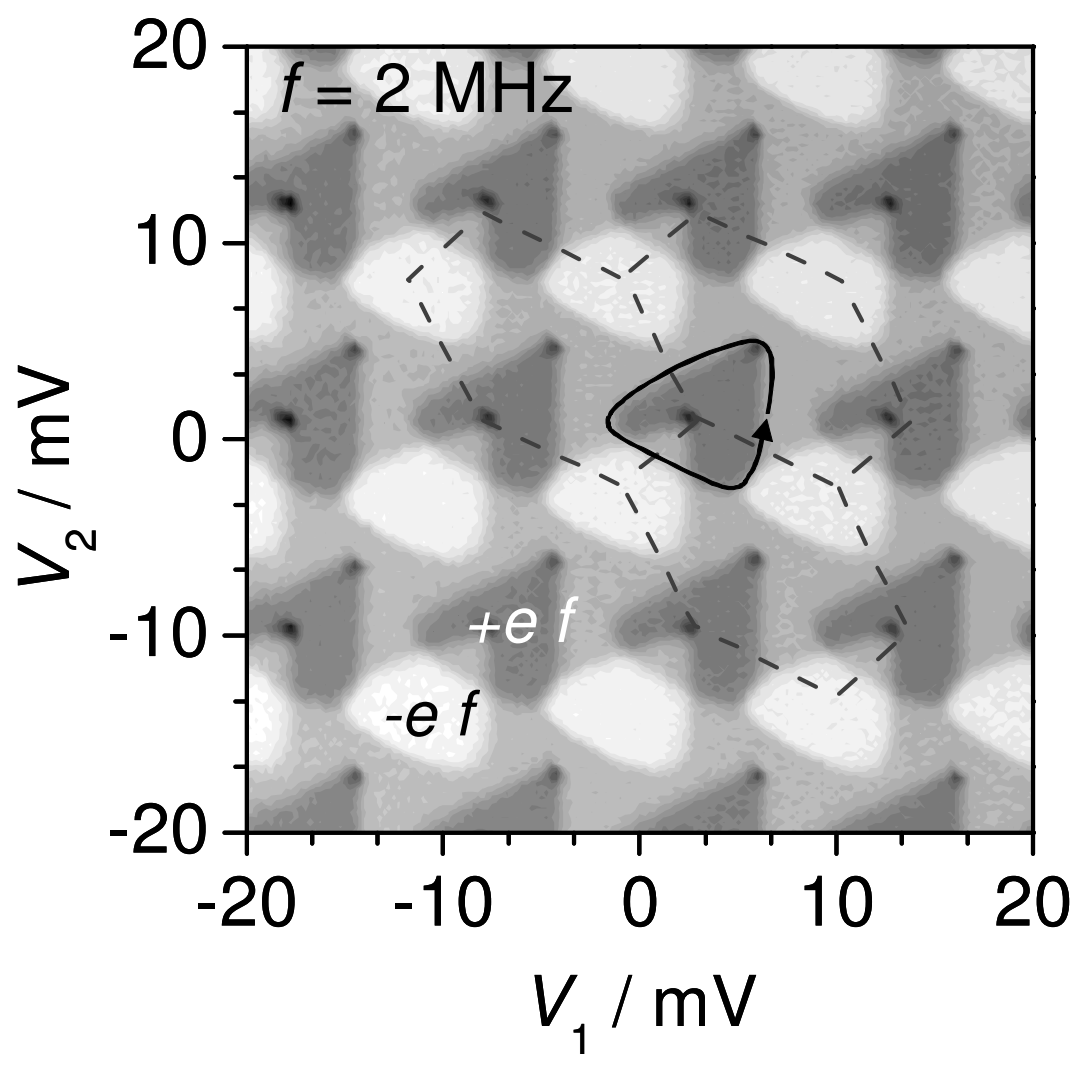

***Fig. 2: Plot of the current (grey-scale) delivered by a 3j r-pump pumping at f = 2 MHz while scanning the dc gate voltages $V_{1,2}$. Within the dark (light) triangular regions, appearing in the typical "honeycomb" pattern, current is pumped in positive (negative) direction. The circumference of these regions (solid curve) reflects the trajectory of the ac gate signals.***

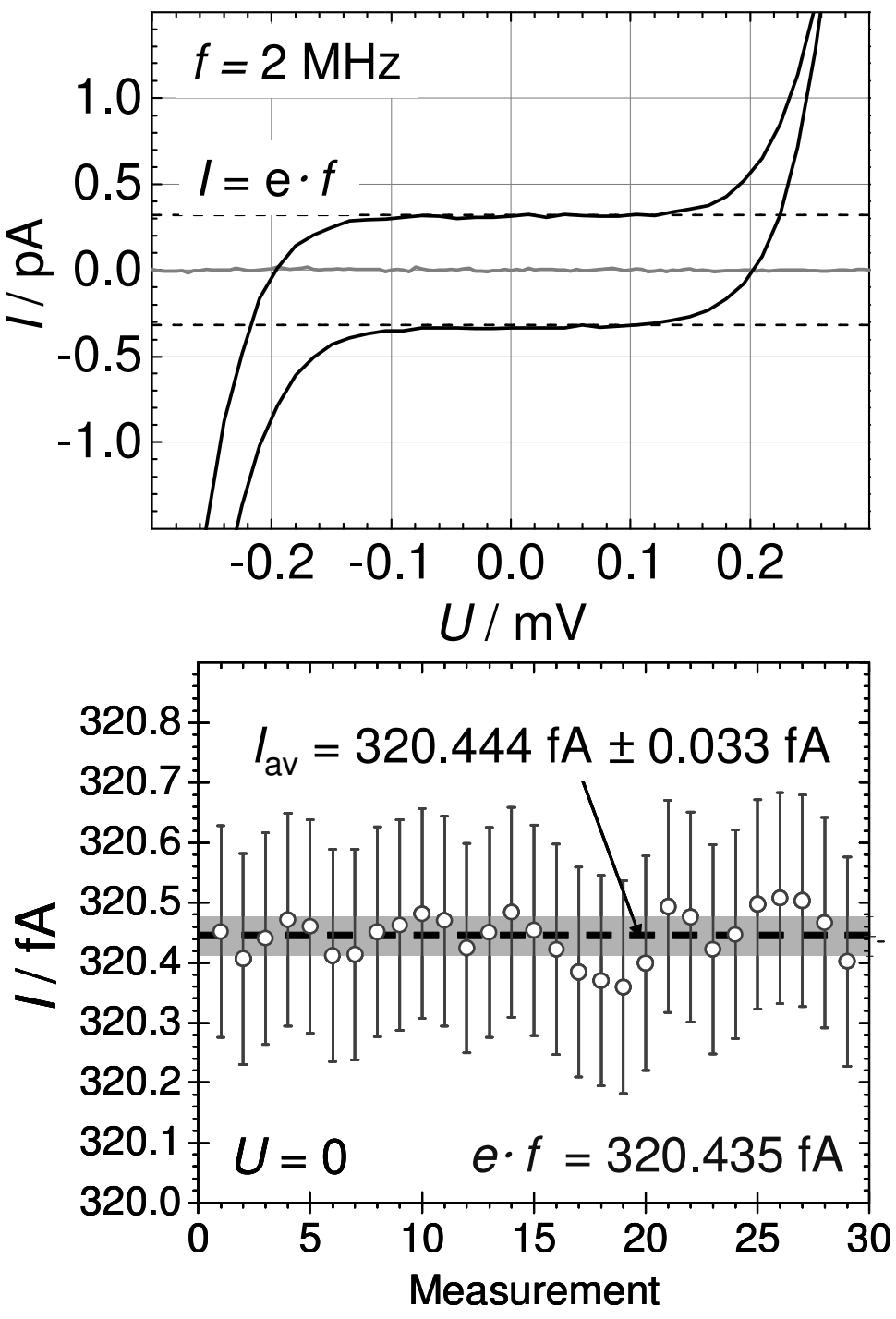

***Fig. 3: I-V curves with current plateaus at $I = ef$ (top panel, $f = 2$ MHz) and precision measurement of the zero-bias current (bottom panel). Each data point was derived from a measurement averaging over two minutes, including current reversal. The grey bar shows the (expanded) uncertainty for the mean value $I_{av}$.***

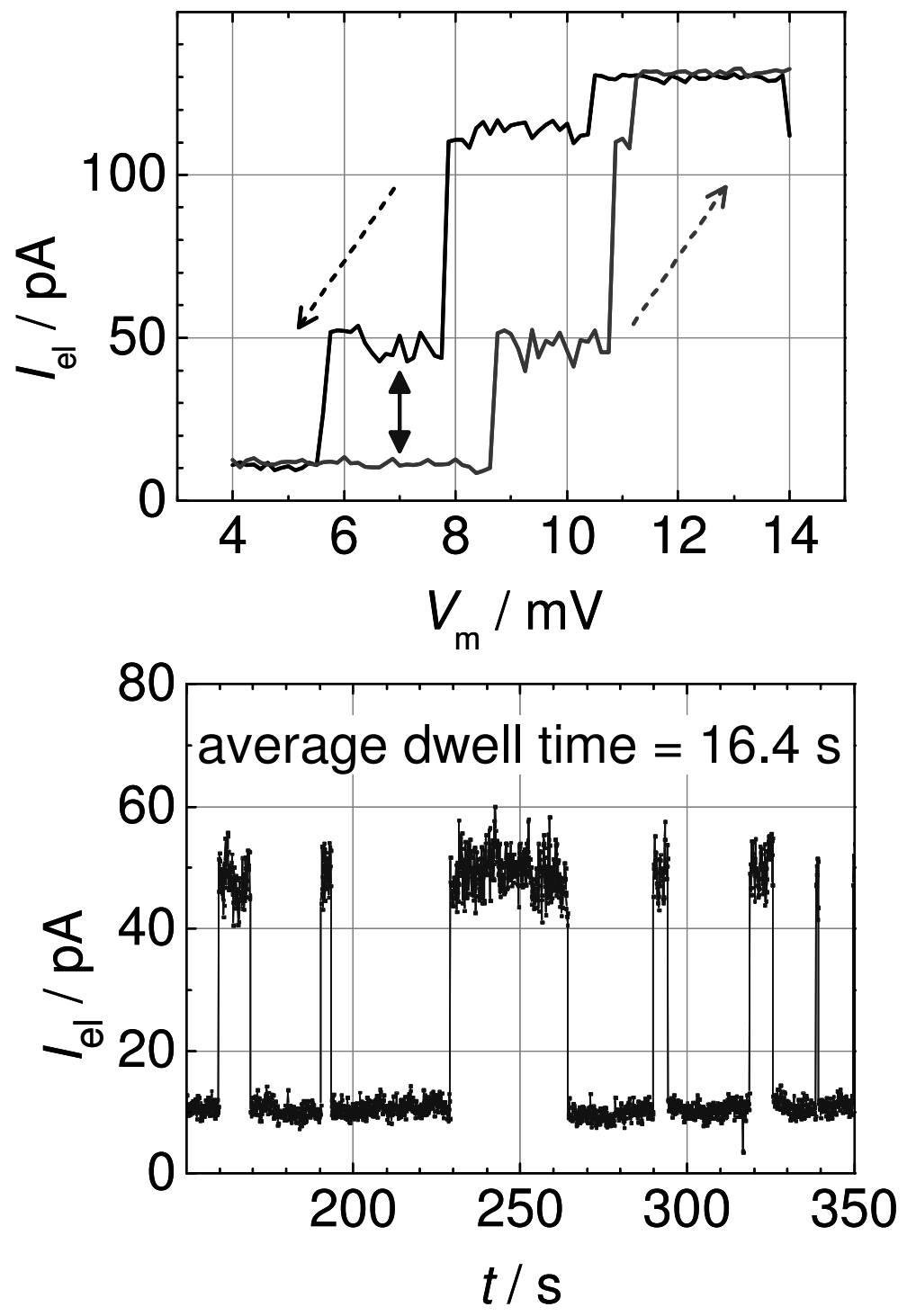

***Fig. 4: Electron trapping effect on type A sample.***

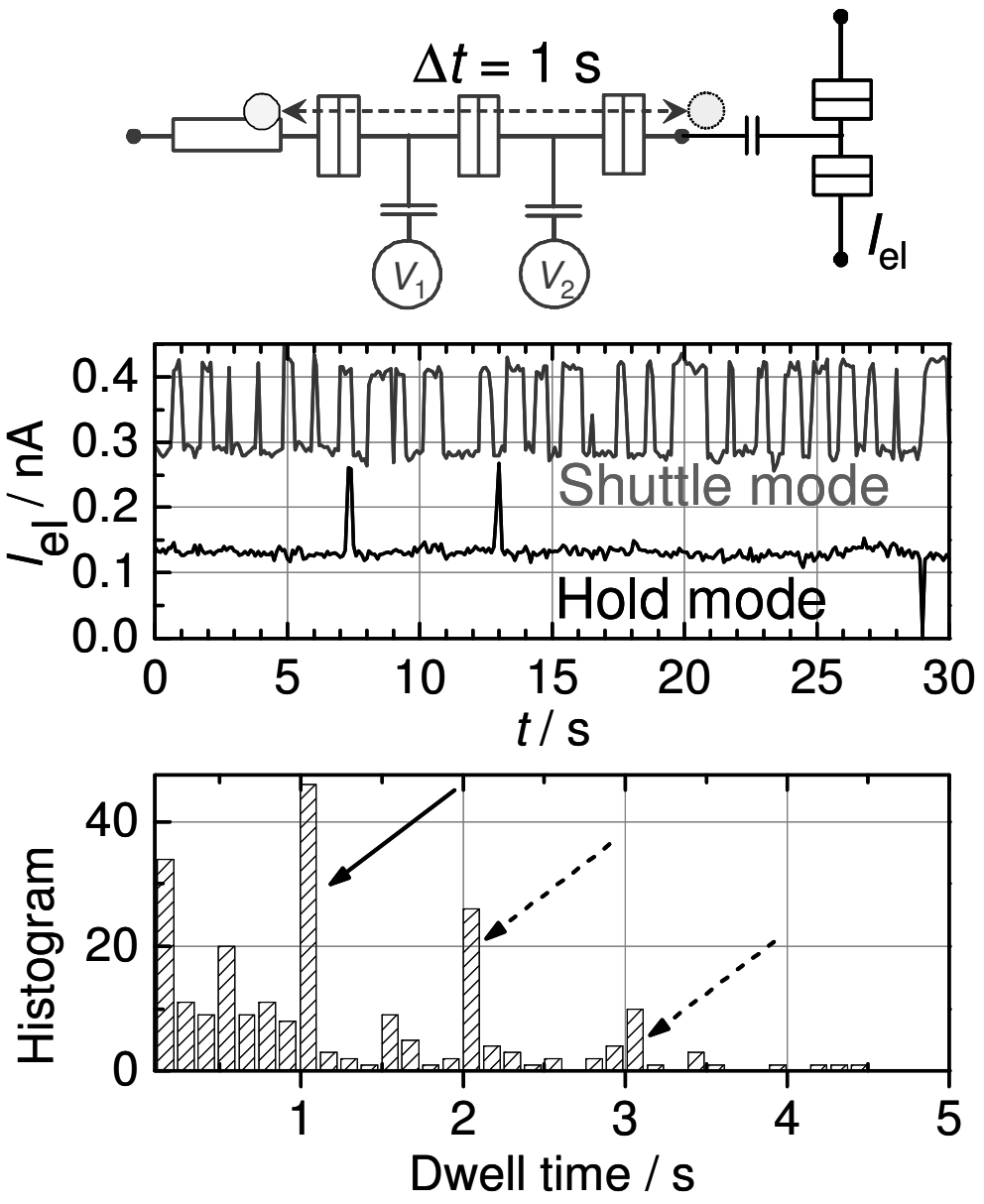

***Fig. 5: Time series traces of the electrometer signal in the hold and the shuttle-pumping mode (±e, wait time 1 s) in type A sample (middle panel, traces are offset for clarity). Because of the small total capacitance only island charge states with + e, 0, and – e were possible. The bottom panel shows a histogram of the electrometer signal during shuttle-pumping.***